\documentclass[preprint,showpacs,singlecolumn,superscriptaddress]{revtex4-1}
\usepackage{graphicx}
\usepackage{tabularx}
\usepackage{color}
\usepackage{amsmath}
\usepackage{comment}
\newcommand{\be}{\begin{equation}}
\newcommand{\ee}{\end{equation}}
\newcommand{\bea}{\begin{eqnarray}}
\newcommand{\eea}{\end{eqnarray}}
\usepackage{dcolumn}
\usepackage{hyperref}
\usepackage{bm}
\usepackage{epsf}
\usepackage{subfigure}
\usepackage{epstopdf}%
\usepackage{amsfonts}

\begin{document}

\title{Energy current and its statistics in the nonequilibrium spin-boson model: Majorana fermion representation}


\author{Bijay Kumar Agarwalla}
\affiliation{Chemical Physics Theory Group, Department of Chemistry, and
Centre for Quantum Information and Quantum Control,
University of Toronto, 80 Saint George St., Toronto, Ontario, Canada M5S 3H6}

\author{Dvira Segal}
\affiliation{Chemical Physics Theory Group, Department of Chemistry, and
Centre for Quantum Information and Quantum Control,
University of Toronto, 80 Saint George St., Toronto, Ontario, Canada M5S 3H6}

\date{\today}
\begin{abstract}
We study the statistics of thermal energy transfer in the nonequilibrium (two-bath) spin-boson model.
This quantum many-body impurity system serves as a canonical model for quantum energy transport.
Our method makes use of the Majorana  fermion representation for the spin operators, in combination with the 
Keldysh nonequilibrium Green's function approach. 
We derive an analytical expression for the cumulant generating function of the model in the steady state limit, and show that
it satisfies the Gallavotti-Cohen fluctuation symmetry. 
We obtain analytical expressions for the heat current and its noise, valid
beyond the sequential and the co-tunnelling regimes. 
Our results satisfy  the quantum mechanical bound for heat current in interacting nanojunctions.  
Results are compared with other approximate theories, as well as with a non-interacting model,
a fully harmonic thermal junction. 
\end{abstract}

\maketitle


\section{Introduction}

The spin-boson (SB) model comprises a two-state system (spin) interacting with a dissipative thermal environment,
a collection of harmonic modes.
It is one of the (conceptually) simplest, yet non-trivial models in the theory of open quantum systems  
\cite{Weiss-book,Leggett-spin}.
The model has found diverse applications in condensed phases physics, chemical dynamics, and quantum optics. In particular,
it offers a rich platform for studying complex physical processes
such as dissipative spin-dynamics \cite{Weiss-book}, charge and energy transfer phenomena in condensed phases 
\cite{Weiss-book, Nitzan-book}, 
Kondo physics \cite{Leggett-spin}, and decoherence dynamics of superconducting qubits \cite{Leggett-spin, Hur-spin}. 
In such applications, the spin system can represent donor-acceptor charge states, a magnetic impurity \cite{Weiss-book},
or a truncated harmonic spectrum, mimicking an anharmonic oscillator \cite{segal-PRL,segal-QME}.
The bosonic bath may stand for a collection of lattice phonons, electromagnetic modes, bound electron-hole pairs, 
and other composite bosonic excitations \cite{Weiss-book,Leggett-spin}.


Beyond questions over quantum decoherence, dissipation, and thermalization, 
which can be addressed by the `canonical' SB model,
the two-bath, nonequilibrium spin-boson (NESB) model has been put forward  as a minimal model for exploring the fundamentals 
of thermal energy transfer in anharmonic nano-junctions \cite{segal-PRL}.
When the two reservoirs are maintained at different temperatures---away from linear response---nonlinear 
functionality such as the diode effect can develop in the junction \cite{segal-PRL,segal-QME, segal-JCP05,WuPRL,ClaireQME}.
More generally, the NESB model serves as a building block for addressing fundamental and practical challenges
in thermal conduction in nanoscale gaps \cite{Saito13,nazim,reviewSA,review-wang,review-li}, 
quantum heat engine operation \cite{segal-pump,segal-stochastic, review-li,berry-1, berry-2},  
molecular conduction junctions \cite{siminePCCP,simine-an,bijay-segal-fcs, reviewSA, Galperin, book-ME} 
and nano-scale energy conversion devices \cite{bijay-segal-fcs,beil}.
%
%


From the theoretical perspective, the NESB model is an extremely rich platform for studying nonequilibrium quantum physics.
One is interested in studying its transport characteristics, including  transient dynamics and steady state properties,
while covering different regimes:
low-to-high temperatures, weak-to-strong system-bath coupling, adiabatic-to-nonadiabatic spin dynamics, 
with or without a (magnetic) spin biasing field, from 
linear response to the far from equilibrium regime.
This challenge could be tackled by extending open quantum system methodologies, 
previously developed to treat the dissipative dynamics of the (traditional) SB model, 
to treat the more complicated, nonequilibrium, two-bath version.

Among the techniques developed to study the characteristics of the thermal heat current in the NESB model we recount
perturbative quantum master equation tools: Redfield equation 
\cite{segal-PRL, segal-JCP05, segal-QME,ClaireQME,Ojanen-QME, Juzar,renJ,Ren}, 
the noninteracting blip approximation (NIBA) \cite{segal-JCP05, segal-QME, segal-Nicolin, Ren},  as well as
Keldysh nonequilibrium Green's function (NEGF) methods \cite{ThossNEGF,yang-EPL}. 
Computational studies  had further established  the non-monotonic behavior of the heat current with 
the spin-bath coupling energy, including studies 
based on the multi-layer multi-configuration Hartree approach \cite{num-multi},
the iterative influence functional path integral technique (in the spin-fermion representation) \cite{num-infi,nazim},
Monte Carlo simulations \cite{Saito13}, and
the hierarchical equation of motion 
\cite{tanimura,brandes}. 


Beyond the analysis of the thermal conductance or the energy current, in small systems
the fluctuations of the current are expected to reveal plethora of information, such as current correlations to all orders
\cite{Esposito-review,Hanggi-review}.
In fact, rather than focusing on the thermal conductance, it is 
more demanding yet highly profitable to pursue
the probability distribution $P_t(Q)$ of the transferred energy $Q$ within a certain interval of time $t$. 
This measure is also known as full-counting statistics (FCS) in the context of electron transport. 
Obtaining the FCS for interacting systems is a highly desirable, yet formidable task.
The FCS of the NESB model has been analyzed so far in two different limits: 
(i) in the sequential tunneling limit i.e., to the lowest order in the system-bath tunnelling strength, 
by employing the Redfield-type quantum master equation approach \cite{renJ,segal-Nicolin},
 and (ii) in the strong coupling and/or high temperature regime, following NIBA-type quantum master equations 
\cite{segal-Nicolin,FCS-strong}. A theory interpolating these two limits was presented in Ref. \cite{Ren}.
However, these studies still miss the low temperature limit \cite{yang-EPL}.

In this paper, we use the Schwinger-Keldysh NEGF approach \cite{Rammer-NEGF,bijay-wang-review} 
in combination with the Majorana fermion representation for the system-spin operators,
 and obtain the cumulant generating function (CGF) of the NESB model beyond the weak spin-bath coupling limit.
The crucial impetus to introduce the Majorana representation 
is that in the fermion representation we are able to use Wick's theorem, 
thus obtain relevant nonequilibrium spin-spin 
correlation functions---while including the counting parameter.
Our results go beyond the sequential and co-tunnelling limits, and we are particularly able to capture to all orders  
the low temperature regime.  As well, we observe deviations from the weak spin-bath coupling limit.
On the other hand, while our result is valid beyond  
the strictly weak-coupling limit, it does miss
the strong-coupling behavior as received in Refs. \cite{num-multi,Saito13,SegalPRE14,nazim,Ren} using the NIBA approach.  
The main outcome of our study is an {\it analytic} expression for the FCS of the NESB model, 
capturing quantum effects, interactions, and far-from-equilibrium function.


The paper is organized as follows.
We introduce the nonequilibrium spin-boson model and the Majorana fermion representation in Sec. \ref{Model}. 
In Sec. \ref{Results}, we  present our main results for the CGF, followed by a discussion over different limits and numerical examples.
We further compare our expressions to previous theories on the NESB model, 
and to the harmonic oscillator-junction model. 
We conclude in Sec. \ref{Summ}. The derivation of the CGF is explained in details in the Appendix.


\section{Model}
\label{Model}

The NESB model comprises a two-state (spin) system coupled to two bosonic reservoirs $(\nu=L,R)$, 
which are maintained at different temperatures. The generic form of the full Hamiltonian is 
\bea
H
= \frac{\hbar \omega_0}{2} \sigma_z + \frac{\hbar \Delta}{2}\sigma_x  + \sum_{j, \nu} \hbar \omega_{j,\nu} b_{j, \nu}^{\dagger} b_{j, \nu}
+  \sigma_z \sum_{j,\nu} \hbar \lambda_{j, \nu }  (b_{j, \nu} + b_{j, \nu}^{\dagger}).
\label{eq:NESB1}
\eea
%
Here, $\sigma_i \,(i=x,y,z)$ are different components of the Pauli matrix, $\omega_0$ and $\Delta$ represents level detuning and the 
hopping between the spin states, respectively. $b_{j, \nu}^{\dagger}$ ($b_{j, \nu}$) is the creation (annihilation) operator of the $j$-th 
phonon mode in the $\nu$-th reservoir. The last term describes the system-bath coupling term with $\lambda_{j, \nu}$ as the coupling strength. 
For simplicity, we focus here on the unbiased case with degenerate spin levels ($\omega_0\!=\!0$). 
Performing a unitary transformation, given by $ U= \frac{1}{\sqrt{2}} (\sigma_x + \sigma_z)$, the transformed Hamiltonian reads
\be
\bar{H}= \frac{\hbar \Delta}{2} \sigma_z  + \sum_{j, \nu} \hbar \omega_{j,\nu} b_{j, \nu}^{\dagger} b_{j, \nu} 
+ {\sigma_x} \sum_{j,\nu} \hbar \lambda_{j, \nu}  (b_{j, \nu} + b_{j, \nu}^{\dagger}).
\ee
%
%
We are interested here in obtaining the steady state energy current and its statistics beyond the weak system-bath coupling limit. 
Unlike the Redfield master equation technique, which captures only resonant energy transfer processes due to its 
underlying weak coupling approximation \cite{segal-QME}, 
the Keldysh nonequilibrium Green's function (NEGF) method  
offers a well established procedure so as to treat the system-bath interaction in a systematic-perturbative way 
\cite{Rammer-NEGF,bijay-wang-review}.
However, the validity of Wick's theorem is a crucial requirement for practicing the method.
Due to the lack of standard bosonic or fermionic commutation relations for spin operators, the NEGF approach is 
in fact {\it unsuitable} to be used in the spin representation of the NESB model. 
However this problem can be avoided by mapping the impurity spin to fermions, using the Majorana-fermion representation \cite{spin1,spin2,yang-EPL}. 


Explicitly, the spin operators can be expressed as $\vec{\sigma}=-\frac{i}{2} \vec{\eta} \times \vec{\eta}$, {\it i.e}., 
\be
\sigma_x = -i \,\eta_y\, \eta_z, \,\, \sigma_y = -i \eta_z \eta_x, \,\, \sigma_z = -i \eta_x \eta_y.
\label{eq:maj1}
\ee
Majorana fermions satisfy the anti commutation relation, 
$\{\eta_{\alpha}, \eta_{\beta}\}=0$, for $\alpha \neq \beta$, 
$\eta_{\alpha}^2=1$, and unlike the Dirac fermions, they are real $\eta_{\alpha}= \eta_{\alpha}^{\dagger}$. 
Therefore,  these fermions can be constructed  in terms of ordinary Dirac fermions ($f$, $g$) 
and their conjugates as  
\bea
\eta_x= (f+f^{\dagger}), \,\,\, \eta_y=i (f^{\dagger}-f), \,\,\,\ \eta_z =(g+g^{\dagger}).
\label{eq:maj2}
\eea 
In this context, it is important to introduce the so-called {\it copy-switching operator} 
\bea
\tau_x = -i \eta_x \eta_y \eta_z,
\label{eq:maj3}
\eea  
in terms of which the Majorana fermions can be expressed as 
$\sigma_{\alpha}= \tau_x \eta_{\alpha}$. 
Note that $\tau_x$ commutes with all Majorana fermion operators and therefore is a constant of motion. 
Also, $\tau_x^2 =1$. With the help of this operator, 
the spin-spin correlator reduces to correlator involving two Majorana fermions
\be
\langle \sigma_{\alpha}(t) \sigma_{\beta}(t')\rangle = \langle \tau_x (t) \eta_{\alpha}(t) \tau_x(t') \eta_{\beta}(t') \rangle = \langle \eta_{\alpha}(t) \eta_{\beta}(t') \rangle.
\label{eq:maj4}
\ee
%
In this mixed Majorana-Dirac representation, the full Hamiltonian reads
\be
\bar{H}= \frac{\hbar \Delta}{2}\,\big(1\!-\! 2 \, f^{\dagger} f\big) +\sum_{j, \nu} \hbar \omega_{j \nu} b_{j, \nu}^{\dagger} b_{j, \nu} + \big(f^{\dagger} \!-\! f\big) \,\eta_z \, (B_L \!+\! B_R),  
\label{full-H}
\ee
where  $B_{\nu}\equiv \sum_{j} \hbar \lambda_{j,\nu} (b_{j, \nu} + b_{j, \nu}^{\dagger})$  
is a $\nu$ bath operator coupled to the spin system.
Note that in this representation, the system-bath coupling term is no longer given in a bilinear form.
For later use, we also identify the components of the Hamiltonian as
$\bar{H} =  H_S + H_L + H_R + H_{SB}$, with
\bea
H_S = \frac{\hbar \Delta}{2}\,\big(1\!-\! 2 \, f^{\dagger} f\big),  \,\,\,\,\,
H_{\nu}= \sum_{j} \hbar \omega_{j, \nu} b_{j, \nu}^{\dagger} b_{j, \nu}, \,\,\,\,\,
H_{SB} = \big(f^{\dagger} \!-\! f\big) \,\eta_z \, (B_L \!+\! B_R).
\eea
%
\section{FCS: Main results}
\label{Results}

\subsection{Working expressions for the FCS}
The complete information over the energy transport statistics can be obtained from the so-called 
cumulant generating function,  ${\cal G}(\xi)$, for heat exchange. 
We begin by defining the energy current operator as the rate of change of energy in one of the reservoirs,
say $L$, and write down the heat current as
$I_L(t)= - \frac{dH^{H}_L(t)}{dt}$. 
The operators are written in the Heisenberg picture, and they evolve with respect to the total Hamiltonian 
$\bar{H}$ in Eq.~(\ref{full-H}). 
Therefore, the total energy change in the $L$ solid within the time interval $t_0=0$ to $t$, where $t_0$ $(t)$ is the initial (final) observation time, 
is given by the integrated current
\be
Q_L(t,t_0)= \int_{t_0=0}^{t} I_L(t') dt' = H_L(0)- H_L^{H}(t).
\ee
Following this definition, we write down the characteristic function ${\cal Z}(\xi)$ 
based on the two-time measurement protocol \cite{Esposito-review, Hanggi-review},
\bea
{\cal Z}(\xi) &=& \Big \langle e^{ i \xi H_L} \, e^{-i \xi H^{H}_L(t)} \Big \rangle = \Big \langle U_{\xi/2}^{\dagger} (t,0)  \, U_{-\xi/2}(t,0)\Big \rangle, \nonumber \\
&=&\Big\langle T_c \exp \Big[-\frac{i}{\hbar} \int_c H^{\xi(\tau)}_{SB}(\tau) d\tau \Big]\Big\rangle,
\label{GF}
\eea
Here, $\xi$ is the ``counting-field'', keeping track of the net amount of energy transferred from the solid $L$ to the spin. 
$\langle... \rangle $ represents an average with respect to the total density matrix at the initial time, $\rho_T(0)$.
We assume a factorized initial state,  $\rho_T(0)=\rho_L(0)\otimes \rho_R(0) \otimes \rho_{S}(0)$, with reservoirs prepared at
a canonical state with inverse temperature $\beta_{\nu}=T_{\nu}^{-1}$,
$\rho_{\nu}(0)=e^{-\beta_{\nu}H_{\nu}}/{\rm Tr_{\nu}}[e^{-\beta_{\nu}H_{\nu}}]$, and an arbitrary state for the spin system $\rho_{S}(0)$.
We also use the definition, 
\bea
U_{p}(t,0) \equiv e^{i p H_L} \, e^{-i {\bar H} t}\, e^{-i p H_L} = e^{-i \bar H_p t/\hbar}, 
\eea
for the counting field-dependent unitary evolution. 
Here, $p = \pm \xi/2$ corresponds to the forward and backward evolution branches. 
Note that due to the measurement protocol, the  modified Hamiltonian $\bar H_p$
acquires a phase in the system-bath coupling term, modifying only the left-bath operators,
%
\be
\bar H_p= \frac{\hbar \Delta}{2}\,\big(1\!-\!2 f^{\dagger} f\big) 
+\sum_{j, \nu} \hbar \omega_{j,\nu} b_{j, \nu}^{\dagger} b_{j, \nu}
+ \big(f^{\dagger} \!-\! f\big)\, \eta_z \,(B^p_L \!+\! B_R)  
\label{eq:mod-Hamiltonian}
\ee
Here,  
$B^p_{L}=\sum_{j} \hbar \lambda_{j,L}(b_{j, L} e^{-i p \hbar \omega_{j, L}} 
+ b_{j, L}^{\dagger} e^{i p \hbar \omega_{j, L}})$  
is the a bath operator,  dressed by the counting field.
In the second line of Eq.~(\ref{GF}), the operators are written in the interaction picture with respect to the 
non-interacting part of the Hamiltonian $H_S + H_L+H_R$. 
$T_c$ is the contour-ordered operator which orders operators according to their contour time;
earlier contour-time operators are placed to the right of later-time terms.
In the long time limit, the CGF is defined as 
\bea
{\cal G}(\xi) \equiv \lim_{t \to \infty} \frac{1}{t}\,\ln {\cal Z}(\xi)=\lim_{t \to \infty} \frac{1}{t}\, \sum_{n=1}^{\infty} \, 
\frac{ (i \xi)^n}{n!}\langle \langle Q^n \rangle \rangle.
\eea
 %
Here,  $\langle \langle Q^n \rangle \rangle$ represent cumulants. Specifically, the second cumulant is
$\langle \langle Q^2 \rangle \rangle = \langle Q^2 \rangle  - \langle Q \rangle^2$.
Taking derivatives of the CGF with respect to $\xi$ immediately hands over the current and its higher order fluctuations, or cumulants. 
However, instead of working with the CGF directly, one can manipulate the so-called generalized current,  defined as
\bea
{\cal I}(\xi)\equiv \frac{\partial {\cal G}(\xi)}{\partial (i \xi)}, 
\eea
by following the nonequilibrium version of Feynman-Hellman theorem first introduced by Gogolin et al. \cite{Gogolin}--- in 
the context of counting statistics for charge transport. 
The key advantage in treating the generalized current, rather than the CGF, lies in the fact that the
problem can be treated with the diagrammatic NEGF technique, as developed originally---without the counting field 
\cite{Schwinger, Kadanoff, Keldysh, book1}.


Using the NEGF with  counting fields as developed in \cite{bijay-Euro}, an expression for 
the generalized energy current can be formally organized as 
\bea
{\cal I}(\xi)\!=\! \int_{-\infty}^{\infty} \frac{d\omega}{4 \pi} \,\hbar \omega \, \Big[\tilde{\Pi}_{xx}^{<}(\omega) \Sigma_L^{>}(\omega) e^{-i \xi \hbar \omega}\!-\! \tilde{\Pi}_{xx}^{>}(\omega) \Sigma_L^{<}(\omega) e^{i \xi \hbar \omega}\Big].
\label{gen-current}
\eea
When  $\xi=0$, this expression reduces to the standard Meir-Wingreen formula \cite{meir-wingreen} for heat current \cite{ThossNEGF}.
The symbol tilde represents that operators within the Green's functions 
evolve with the dressed (counting field-dependent) Hamiltonian $\bar H_p$ given in Eq.~(\ref{eq:mod-Hamiltonian}). 
$\tilde{\Pi}_{xx}^{<, >}(\omega)$ are the Fourier transformed lesser and greater components of the spin-spin correlators, 
namely, 
\bea
&&\tilde{\Pi}^{<}_{xx}(t,t') = -i \langle \sigma_x (t') \sigma_x(t)\rangle_{\xi}
\nonumber\\
&& \tilde{\Pi}^{>}_{xx}(t,t') = -i \langle \sigma_x (t) \sigma_x(t')\rangle_{\xi}.
\eea
%
$\Sigma_{\nu}^{<,>}(\omega)$ are the self-energy components emerging due to the coupling of the spin to the solids, 
responsible for transferring energy in and out of the system,
\bea
&& \Sigma_{\nu}^{<}(\omega) = - i \, n_{\nu}(\omega) \, \Gamma_{\nu}(\omega)
\nonumber\\
&& \Sigma_{\nu}^{>}(\omega) = - i \,   \bar{n}_{\nu}(\omega) \, \Gamma_{\nu}(\omega).
\eea  
Here,  $\bar{n}_{\nu}(\omega)\equiv [1+ n_{\nu}(\omega)]$ with $n_{\nu}(\omega)= (e^{\beta_{\nu} \hbar \omega}-1)^{-1}$ 
as the Bose-Einstein distribution function and $\beta_{\nu}= 1/T_{\nu}$ is the inverse temperature. 
$\Gamma_{\nu}(\omega) = {2 \pi} \sum_{j} \lambda_{j, \nu}^2 \, \delta(\omega\!-\!\omega_j)$ 
is the spectral function for the $\nu$ reservoir.

Note that we write integrals covering negative frequencies, by extending the range of the spectral function while
satisfying $\Gamma_{\nu}(\omega)=-\Gamma_{\nu}(-\omega)$. 

\subsection{Main results}

To receive the generalized current, our primary objective is to obtain the components 
$\tilde{\Pi}_{xx}^{</>}(\omega)$.
These terms are obtained using the NEGF method following a first order perturbation expansion
with respect to the interaction of the bath with the spin. 
We summarize here the central results; details are given in the Appendix.

The lesser and greater components are obtained to the lowest non-zero order in the nonlinear self-energy. They are given as 
\bea
\tilde{\Pi}_{xx}^{<}(\omega) 
&=& -\frac{4 i \Delta^2}{{\cal D}(\omega,\xi)} \Big( \Gamma_L(\omega)  n_L(\omega) e^{i \xi \hbar \omega} + \Gamma_R(\omega) n_R(\omega) \Big), \\
\tilde{\Pi}_{xx}^{>}(\omega)
&=& -\frac{4 i \Delta^2}{{\cal D}(\omega,\xi)} \Big( \Gamma_L(\omega)  \bar{n}_L(\omega) e^{-i \xi \hbar \omega} + \Gamma_R(\omega) \bar{n}_R(\omega) \Big),
\eea
with 
\be
{\cal D}(\omega, \xi)= (\omega^2 -\Delta^2)^2 + \omega^2  M(\omega,\xi).
\ee
Here 
$M(\omega,\xi)=C^2(\omega) + 4 \, A(\omega,\xi)$ includes the two terms,
\bea
C(\omega) &=& \Gamma_L (\omega) \left[1+ 2 n_L (\omega)\right] 
+ \Gamma_R(\omega) \left[1+ 2 n_R (\omega)\right], \nonumber \\ 
{A}(\omega,\xi) &=& \Gamma_L(\omega) \Gamma_R(\omega) 
\Big [ n_L(\omega) \, {\bar n}_R(\omega) (e^{i \xi \hbar \omega}\!-\!1) + n_R(\omega) \, \bar{n}_L(\omega) (e^{-i \xi \hbar \omega}\!-\!1) \Big].
\label{eq:CA}
\eea
If we eliminate the counting parameter, $\xi=0$,  $\Pi_{xx}^{</>}(\omega)$ provides the imaginary components of the response function 
$\Pi_{xx}^{R}(\omega)$,
\be
Im[\Pi_{xx}^R(\omega)]= -\frac{2\, \Delta^2 \,\big(\Gamma_L(\omega)+\Gamma_L(\omega)\big)}{(\omega^2 -\Delta^2)^2 
+ \omega^2 \big[\Gamma_L (\omega) (1+ 2 n_L (\omega)) + \Gamma_R(\omega) (1+ 2 n_R (\omega))\big]^2},
\ee 
matching the results of Ref.~(\onlinecite{yang-EPL}).

Using these expressions, the CGF for the NESB model, ${\cal G}_{SB}(\xi) \equiv \int_{0}^{\xi} d\xi' \, {\cal I}(\xi')$, 
is obtained as
\bea
{\cal G}_{SB}(\xi) \!&\!=\!\!&\int_{-\infty}^{\infty}\! \frac{d\omega}{4 \pi}\, \frac{\Delta^2}{\omega^2}\, \! \ln  
\Big\{ 1 + {\cal T}_{SB}(\omega; T_L, T_R) \Big[ n_L(\omega)\bar{n}_R(\omega)  
\nonumber \\
&&( e^{i \xi \hbar \omega}\!-\!1) + n_R(\omega)\bar{n}_L(\omega) (e^{-i \xi \hbar \omega}\!-\!1)\Big] \Big\},
\label{eq:CGF-central}
\eea
with the temperature-dependent transmission function
\be
{\cal T}_{SB}(\omega; T_L,T_R) = \frac{4 \,\Gamma_L(\omega) \,\Gamma_R(\omega) \,\omega^2}{ \Big(\omega^2 \!-\!\Delta^2\Big)^2 + \omega^2 \, \Big[\Gamma_L(\omega) (1\!+\! 2 n_L(\omega)) + \Gamma_R(\omega) (1\!+\! 2 n_R(\omega))\Big]^2}.
\label{eq:SB-trans}
\ee
This expression is valid with an arbitrary form for the spectral function $\Gamma_{\nu}(\omega)$. 
The CGF further satisfies the steady state Gallavotti-Cohen fluctuation symmetry, 
${\cal G}(\xi)= {\cal G}(-\xi+ i (\beta_R-\beta_L))$ \cite{GC-sym}.
Eq. (\ref{eq:CGF-central}) constitutes the main result of our work.

The cumulants of the  energy flux can be readily obtained
by taking derivatives of the CGF with respect to the counting field $\xi$. 
For example, the heat current and its noise are given by
\be
\langle I \rangle_{SB}\!  \equiv \! \frac{\partial {\cal G}_{SB}(\xi)}{\partial (i\xi)}\Big{|}_{\xi=0} = 
\int_{-\infty}^{\infty} \frac{d\omega}{4 \pi} \,\frac{\hbar\Delta^2} { \omega}\, {\cal T}_{SB}(\omega;T_L,T_R) \,\big[n_L(\omega)-n_R(\omega)\big],
\label{eq:SB-current}
\ee
\bea
\langle S \rangle_{SB} \!& \equiv& \!\frac{\partial^2 {\cal G}_{SB}(\xi)}{\partial (i\xi)^2}\Big{|}_{\xi=0} 
= \int_{-\infty}^{\infty} \frac{d\omega}{4 \pi} (\hbar \Delta)^2 \, \Big\{-{\cal T}_{SB}^2(\omega;T_L,T_R) \big[n_L(\omega)-n_R(\omega)\big]^2 \nonumber \\
&+& {\cal T}_{SB}(\omega;T_L,T_R)  \big[n_L(\omega) \bar{n}_R(\omega) \!+\! n_R(\omega) \bar{n}_L(\omega)\big] \Big\}.
\label{cur-noise}
\eea
The result for the current agrees with the derivation in Ref. \cite{yang-EPL}---once
we organize  our expressions,  $ \int_{-\infty}^{\infty} d\omega ... \rightarrow 2 \times \int_0^{\infty} d\omega$...
In the next subsection, we discuss interesting limits of the general results.

\subsection{Special limits}

\noindent
{\it Incoherent sequential tunnelling.} When the system-bath coupling is weak 
and the reservoirs' temperatures are high, $ \Gamma_{L,R} \ll \Delta \leq T_{L,R}$,  the above generating function reduces to the result  
obtained from the Redfield quantum master equation approach \cite{segal-Nicolin}, 
when directly employing the Born-Markov approximation. 
We now derive this result.  
Following Eq.~(\ref{eq:CGF-central}), the generalized current can be simplified to 
\be
{\cal I}_{SB}(\xi) = \int_{-\infty}^{\infty} \, \frac{d\omega}{4\pi} \frac{\,\Delta^2 \, M'(\omega,\xi)}{(\omega^2 \!-\!\Delta^2)^2 \!+\! \omega^2 M(\omega,\xi)},
\label{eq:currweak}
\ee 
where $M'(\omega,\xi) = \frac{\partial M(\omega,\xi)}{\partial{(i \xi)}}$. 
To the lowest order $O(\Gamma_{L,R}^2)$, working in the limit  $ \Gamma_{L,R} \ll \Delta \leq T_{L,R}$,
the poles in the integrand can be approximated by 
\be
\pm \Big\{\Delta \pm \frac{i}{2} \sqrt {M(\Delta,\xi)}\Big\}.
\ee
By employing the residue theorem, the integration in Eq. (\ref{eq:currweak}) results in 
${\cal I}_{SB}^{weak}(\xi) =\frac{1}{2} \frac{\partial \sqrt{M(\xi)}}{\partial (i\xi)}$ and the generating function reduces to  
\be
{\cal G}_{SB}^{weak}(\xi) = -\frac{1}{2} \Big(C(\Delta)\! -\!\sqrt{ M(\Delta,\xi)}\Big).
\label{CGF-weak}
\ee
This expression matches  the result obtained in Ref.~\cite{segal-Nicolin}. This CGF also respects the fluctuation symmetry. 
It immediately yields the heat current in the weak coupling limit \cite{segal-PRL}
\bea
\langle I\rangle_{SB}^{weak} = 
\hbar \Delta \frac{\Gamma_L(\Delta)\Gamma_R(\Delta)
\left[ n_L(\Delta)-n_R(\Delta)\right]}
{\left[\Gamma_L(\Delta)(1+2n_L(\Delta))\right] + \left[\Gamma_R(\Delta)(1+2n_R(\Delta))\right]}.
\label{eq:ISBweak}
\eea
%


\noindent
{\it Co-tunnelling.} 
At low temperatures, $\Gamma_{\nu}\ll T_{\nu}\leq \Delta$,  
the process of sequential tunnelling is exponentially suppressed 
since incoming phonons are off-resonance---with frequencies below the spin energy gap, $\omega \ll \Delta$.
The dominant contribution to the current and higher order fluctuations thus
comes from coherent two-phonon co-tunnelling processes. In this limit, 
the transmission function of Eq. (\ref{eq:SB-trans}) 
is given by ${\cal T}_{SB}^{co}(\omega,T_L,T_R) \sim 4 \Gamma_L(\omega) \Gamma_R(\omega) \omega^2/ \Delta^4
\ll1$. By approximating $\ln(1+x)\sim x$ for small $x$, we reduce the CGF of Eq. (\ref{eq:CGF-central}) to
\be
{\cal G}_{SB}^{co}(\xi)= \frac{2}{\pi} \int_{0}^{\omega_h} {d\omega} \, \frac{\Gamma_L(\omega) \Gamma_R(\omega)}{\Delta^2} \, \Big( n_L(\omega) \bar{n}_R(\omega) ( e^{i \xi \hbar \omega}\!-\!1) + n_R(\omega) \bar{n}_L(\omega) (e^{-i \xi \hbar \omega}\!-\!1)\Big),
\label{CGF-cotunn}
\ee
with fluctuation symmetry being satisfied. Here, $\omega_h$, the upper limit in the integral should be determined by the
smaller energy scale, temperature of the cutoff frequency of the baths.
The co-tunneling (co) heat current then becomes 
\bea
\langle I \rangle_{SB}^{co} = \frac{2}{\pi} \int_0 ^{\omega_h} d\omega 
 \hbar\omega \frac{\Gamma_L(\omega) \Gamma_R(\omega) }{\Delta^2} 
\left[ n_L(\omega)-n_R(\omega)\right].
\label{eq:ISBco}
\eea
%
This expression was previously achieved in two ways:
(i) By using a systematic perturbative treatment \cite{Ojanen-QME}, and (ii)
working with the so-called Born-Oppenheimer approach
for heat exchange \cite{BO}, by assuming slow bath and a fast (high frequency) impurity.
In the case of an Ohmic bath, $\Gamma_{\nu}(\omega)\propto \omega^s$ with $s=1$,
the heat current scales as
$\langle I \rangle_{SB}^{co} \propto T_L^4-T_R^4$, thus the thermal conductance scales with $T^3$, 
in agreement with numerically exact simulations on the NESB model \cite{Saito13}.
As well, in this low temperature limit the NESB junction behaves similarly to a fully harmonic junction, as we discuss in
Sec. \ref{SecH}.

Note that in contrast to the CGF received in Eq.~(\ref{eq:CGF-central}) and Eq.~(\ref{CGF-weak}), 
the CGF in the co-tunnelling limit is {\it symmetric} with respect to $\Gamma_{L,R}(\omega)$. Therefore, in this limit
the system does not support the thermal rectification effect. 
Moreover, in this limit the cumulants $C^n = \frac{\partial^n{\cal G}_{SB}(\xi)}{\partial (i\xi)^n}\big{|}_{\xi=0}$ 
scale as $C^n \propto 1/\Delta^2$, whereas in the sequential tunneling limit cumulants grow 
as $C^n \propto \Delta^n$. 

\subsection{Comparison between the NESB model and the harmonic oscillator junction}
\label{SecH}

In the harmonic oscillator (HO) junction, a single harmonic oscillator of frequency $\omega_0$,
replaces the spin impurity of the NESB model, Eq. (\ref{eq:NESB1}).
The resulting Hamiltonian is fully harmonic, and it can be 
readily solved exactly to yield the CGF  \cite{saito-fcs,bijay-PRE-fcs}
\bea
{\cal G}_{HO}(\xi)\! &\!=-&\int_{-\infty}^{\infty}\! \frac{d\omega}{4 \pi}\, \! \ln  \Big[ 1 - {\cal T}_{HO}(\omega) \Big(n_L(\omega) \bar{n}_R(\omega)  \nonumber \\
&&( e^{i \xi \hbar \omega}\!-\!1) + n_R(\omega) \bar{n}_L(\omega) (e^{-i \xi \hbar \omega}\!-\!1)\Big) \Big].
\eea
Surprisingly, our final expression for the CGF of the NESB model,  Eq.~(\ref{eq:CGF-central}), 
is very similar to this expression. The following differences show up: (i) In the HO case the transmission function does not
depend on the temperatures of the baths, 
\be
{\cal T}_{HO}(\omega)= \frac{4 \,\Gamma_L(\omega) \,\Gamma_R(\omega) \,\omega^2}{ \Big(\omega^2 \!-\!\omega_0^2\Big)^2 + \omega^2 \, \Big(\Gamma_L(\omega) + \Gamma_R(\omega)\Big)^2}.
\label{eq:THO}
\ee
Further, (ii)  there is  a crucial sign difference in this CGF as compared to $\mathcal{G}_{SB}(\xi)$ in Eq.~(\ref{eq:CGF-central}).  This sign difference reflects on  the nonlinear nature of the spin. A similar sign-difference between harmonic and spin impurity nanojunctions
has been observed in vibrationally-assisted  electron conducting junctions \cite{beil,bijay-segal-fcs}.
The above expression immediately provides the Landauer expression for the heat current,
\bea
\langle I \rangle_{HO} = \frac{1}{4\pi}\int_{-\infty}^{\infty}d\omega \hbar \omega {\cal T}_{HO}(\omega)[n_L(\omega)-n_R(\omega)],
\label{eq:IHO}
\eea
%
and the noise 
\bea
\langle S \rangle_{HO} = \frac{1}{4\pi}\int_{-\infty}^{\infty}d\omega 
(\hbar\omega)^2 \left[ {\cal T}_{HO}^2(\omega)
\left( n_L(\omega)-n_R(\omega)\right)^2 + {\cal T}_{HO}(\omega)\left(n_L(\omega)\bar n_R(\omega)+ \bar n_R(\omega)n_L(\omega) \right)\right].
\nonumber\\
\label{eq:SHO}
\eea
In the  weak coupling limit, the CGF of the HO model reduces to the standard result obtained by a low order QME \cite{renJ,segal-Nicolin}  
\bea
{\cal G}_{HO}^{weak}(\xi) &=& \frac{1}{2} \Big(C_{HO}(\omega_0)\!-\!\sqrt{C_{HO}^2(\omega_0)\!-\!4\, A_{HO}(\omega_0, \xi)}\Big),
\eea
with 
\bea
C_{HO}(\omega_0) &=& \Gamma_L(\omega_0) \!+\! \Gamma_R(\omega_0), \nonumber \\
A_{HO}(\omega_0, \xi) &=& \Gamma_L(\omega_0) \Gamma_R(\omega_0) \Big ( n_L(\omega_0) \, {\bar n}_R(\omega_0) (e^{i \xi \hbar \omega_0}\!-\!1) + n_R(\omega_0) \, \bar{n}_L(\omega_0) (e^{-i \xi \hbar \omega_0}\!-\!1) \Big). \nonumber 
\eea
The heat current then reduces to the familiar result,
\bea
\langle I \rangle_{HO}^{weak} = \hbar \omega_0 \frac{\Gamma_L(\omega_0)\Gamma_R(\omega_0)}{\Gamma_L(\omega_0)+\Gamma_R(\omega_0)}
\left[n_L(\omega_0)-n_R(\omega_0) \right].
\eea
%
%
The co-tunnelling limit is more subtle, and we exemplify it now when calculating the current. 
We break the transmission function (\ref{eq:THO}) into two contributions (leaving for a moment the numerator) 
${\cal T}_{HO}(\omega)= {\cal T}_{o}(\omega) + {\cal T}_{e}(\omega)$,
%
\bea
{\cal T}_{o}(\omega) &=&
 \frac{\omega^2-\omega_0^2}{(\omega^2-\omega_0^2)^2+[\Gamma_L(\omega)+\Gamma_R(\omega)]^2\omega^2}
\nonumber\\
{\cal T}_{e}(\omega)&=&
 \frac{\omega_0^2}{(\omega^2-\omega_0^2)^2+[\Gamma_L(\omega)+\Gamma_R(\omega)]^2\omega^2 }. 
\eea
Assuming the hierarchy of energies $\Gamma_{\nu}\ll T_{\nu}\leq \omega_0$, 
we note that the function $\omega[n_L(\omega)-n_R(\omega)]$
changes slowly at the vicinity of $\omega_0$, in the regime where the functions ${\cal T}_{e,o}(\omega)$ have significant weight.
Therefore, the integral (\ref{eq:IHO}) over the odd component (approximately) cancels out, and the current is solely determined
by the even term, ${\cal T}_e(\omega)\sim 1/\omega_0^2$, to yield
%
 \bea
\langle I \rangle _{HO}^{co}=\frac{2}{\pi}\int_{0}^{\omega_h} d\omega \hbar \omega\frac{\Gamma_L(\omega)\Gamma_R(\omega)}{\omega_0^2}
[n_L(\omega)-n_R(\omega)].
\label{IHOco}
\eea
This result reproduces exactly the behavior of the NESB model in the corresponding limit, Eq. (\ref{eq:ISBco}).
This correspondence is not surprising: At low temperatures (smaller than the energy spacing in the quantum impurity) and at
weak system-bath coupling,  the NESB and the HO junctions should behave rather similarly. 
For a comprehensive analysis of the harmonic-mode thermal junction, see Ref. \cite{Andrei}.

\subsection{Steady state population and a bound on heat current}

Besides transport properties, we use
the Majorana formalism and calculate the steady state population of the
ground and excited states in the eigenbasis of the spin. 
This can be obtained by calculating $\langle \sigma_z\rangle$, given as, 
\bea
\langle \sigma_z \rangle &=& i  \int_{-\infty}^{\infty} \frac{d\omega}{4 \pi} \, \Big[\left(\begin{array}{cc} 1 & 1 \end{array}\right ) {G}_{\Psi}^{>}(\omega) \left(\begin{array}{c}1 \\-1 \end{array}\right) - \left(\begin{array}{cc} -1 & 1 \end{array}\right ) {G}_{\Psi}^{<}(\omega) \left(\begin{array}{c}1 \\1 \end{array}\right) \Big] \nonumber \\
&=& -\frac{\Delta}{\pi} \int_{-\infty}^{\infty} {d\omega} \, \omega \, \frac{\Gamma_L(\omega)  + \Gamma_R(\omega)}{ (\omega^2-\Delta^2)^2 + \omega^2\,C^2(\omega)}.\quad
\eea
The function $C(\omega)$ is defined in Eq. (\ref{eq:CA}).
In the weak coupling limit, we receive the same result as obtained in Ref.~\cite{segal-QME},
\be
\langle \sigma_z \rangle_{weak} = - \frac{\Gamma_L(\Delta) + \Gamma_R(\Delta)}
{\Gamma_L(\Delta)  \left[1+2{n}_L(\Delta)\right]+\Gamma_R(\Delta) \left[1+2{n}_R(\Delta)\right]}.
\ee
The population of the states are 
$p_g= \frac{1}{2} (1 -\langle \sigma_z \rangle)$ and $p_e= \frac{1}{2} (1 +\langle \sigma_z \rangle)$.

Recently, a rigorous quantum mechanical bound for the heat current in interacting systems has been derived, valid
at the high temperature---yet in the quantum regime \cite{Ed-bound}. 
We now confirm that the heat current derived in our work, Eq.~(\ref{eq:SB-current}),
does not violate the bound. This further affirms the validity and usefulness of our result.

In the following analysis we make use of 
the inequality $0 \leq [n_L(\omega)-n_R(\omega)] \leq (T_L-T_R)/(\hbar \omega)$ for $\omega >0$ and $T_L >T_R$. As well, we
recall on the positivity of the transmission function ${\cal T}_{SB}(\omega)>0$. Furthermore,
we assume an Ohmic spectral density function for the reservoirs, 
$\Gamma_{\nu}(\omega) = \gamma_{\nu} \,\omega, \nu =L,R$ (see Ref. \cite{Ed-bound} for a detailed discussion over different spectral functions).
Putting these pieces together, we conclude that the heat current of Eq. (\ref{eq:SB-current}) satisfies the following inequality
\bea
\langle I\rangle &\leq&\int_{-\infty}^{\infty} \frac{d\omega}{4 \pi} \,\, \frac{ 4 \gamma_L \gamma_R \Delta^2 \omega^2 (T_L-T_R)}{(\omega^2-\Delta^2)^2 + \omega^2 C^2(\omega)} \nonumber \\
&=& \frac{2}{\pi} \,\Delta^2 (T_L-T_R) \frac{\gamma_L \gamma_R}{\gamma_L+\gamma_R} \, \int_{0}^{\infty} d\omega \, \omega^2 \frac{\gamma_L + \gamma_R}{(\omega^2-\Delta^2)^2 + \omega^2 C^2(\omega)} \nonumber \\
& =& -\Delta  (T_L-T_R) \frac{\gamma_L \gamma_R}{\gamma_L+\gamma_R}  \langle \sigma_z \rangle
\eea
%
which precisely matches with the bound organized in Ref.~\cite{Ed-bound} for the NESB model.
We conclude that our expression for the current thus does not violate a fundamental bound, unlike the prediction of the Redfield QME, see
Ref.~\cite{Ed-bound}.


\section{Numerical Results}
\label{eq:simul}

\begin{figure}
\includegraphics[width=18cm]{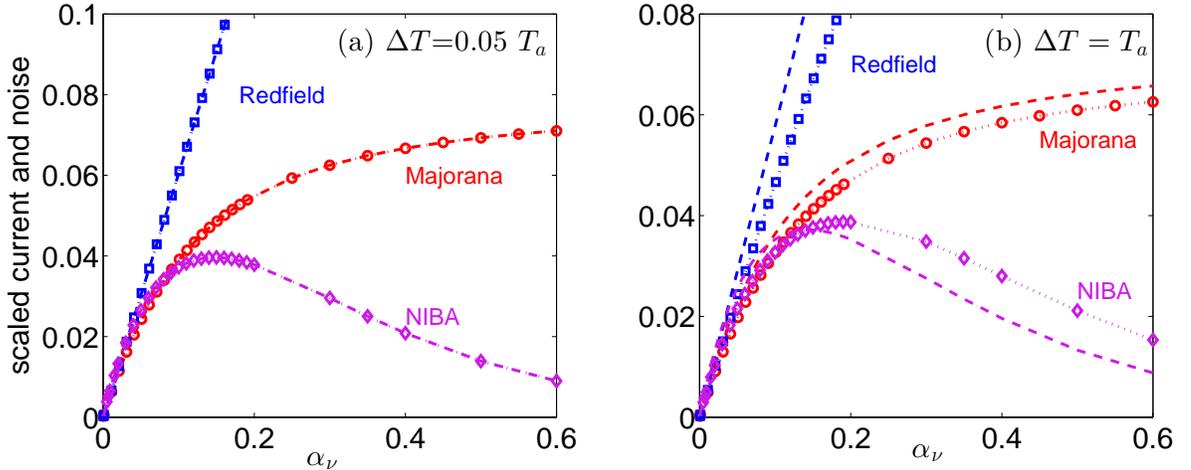}
\caption{ 
Scaled current $\langle I \rangle /\Delta T$ (dashed lines) 
and noise $\langle S \rangle /2 T_a$ (symbols) for the NESB model as
a function of coupling strength $\alpha_\nu$, employing different
theoretical schemes: Redfield (blue), Majorana (Red) and NIBA (purple).
(a) Results close to equilibrium, $\Delta T=0.05 T_a$. 
(b) Calculations far-from-equilibrium, $\Delta T=T_a$, demonstrating deviations from
the fluctuation-dissipation theorem.
Parameters are $\Delta=T_a=1$, $T_{L,R}=T_a\pm\Delta T/2$, $\omega_c=10\Delta$, and 
$\alpha_{L}=\alpha_R$.
}
\label{Fig1}
\end{figure}

\begin{figure}
\includegraphics[width=16cm]{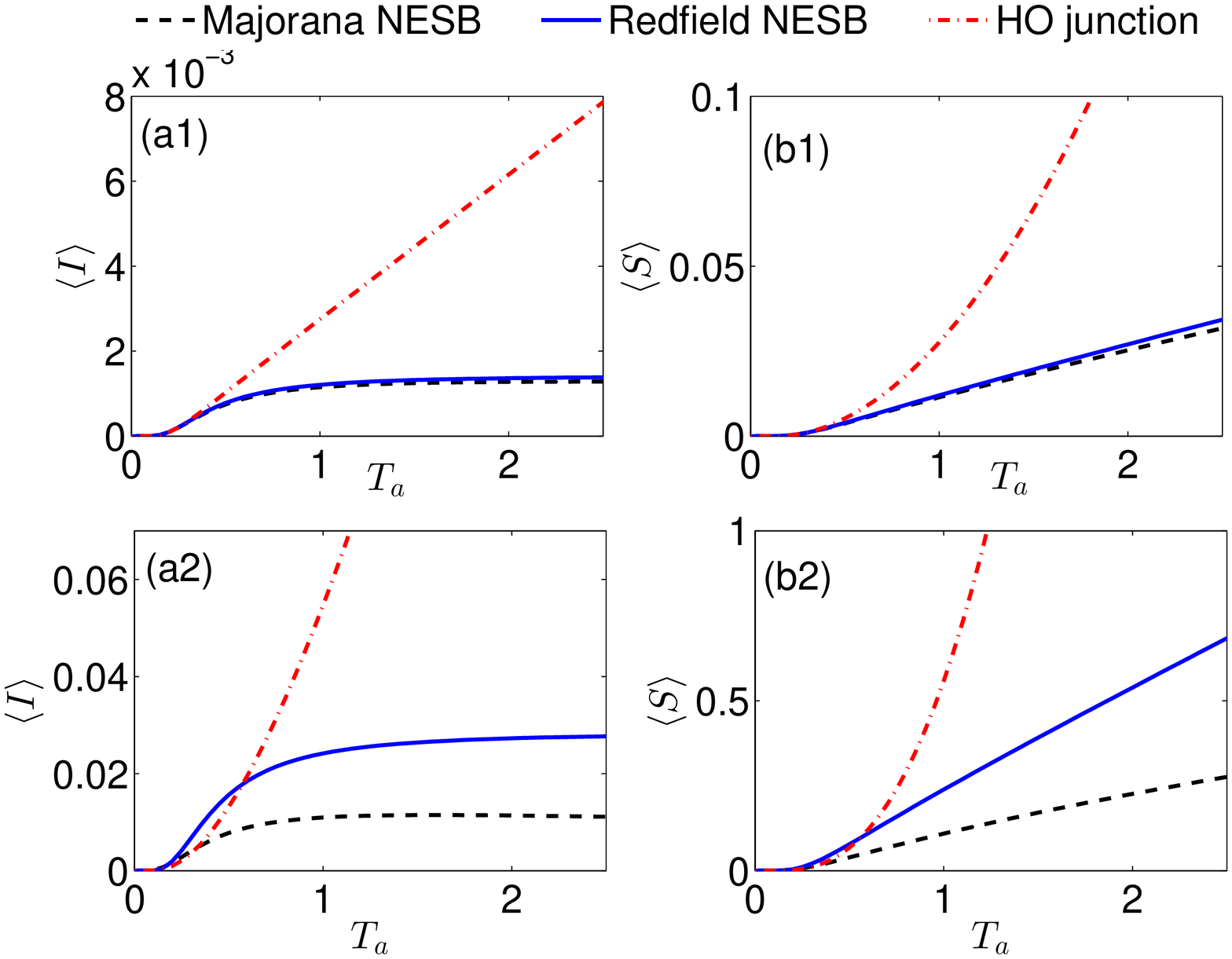}
\caption{ 
Temperature dependence of the heat current and  noise
for the SB junction (Redfield and Majorana) and the HO model.
(a1)-(b1) Weak coupling limit $\alpha_{\nu}=0.01$.
(a2)-(b2) Intermediate coupling, $\alpha_{\nu}=0.2$.
Parameters are $\Delta=1$, $\Delta T=0.2 T_a$ and
$T_{L,R}=T_a\pm\Delta T/2$, $\omega_c=10\Delta$.}
\label{Fig2}
\end{figure}

\begin{figure}
\includegraphics[width=18cm]{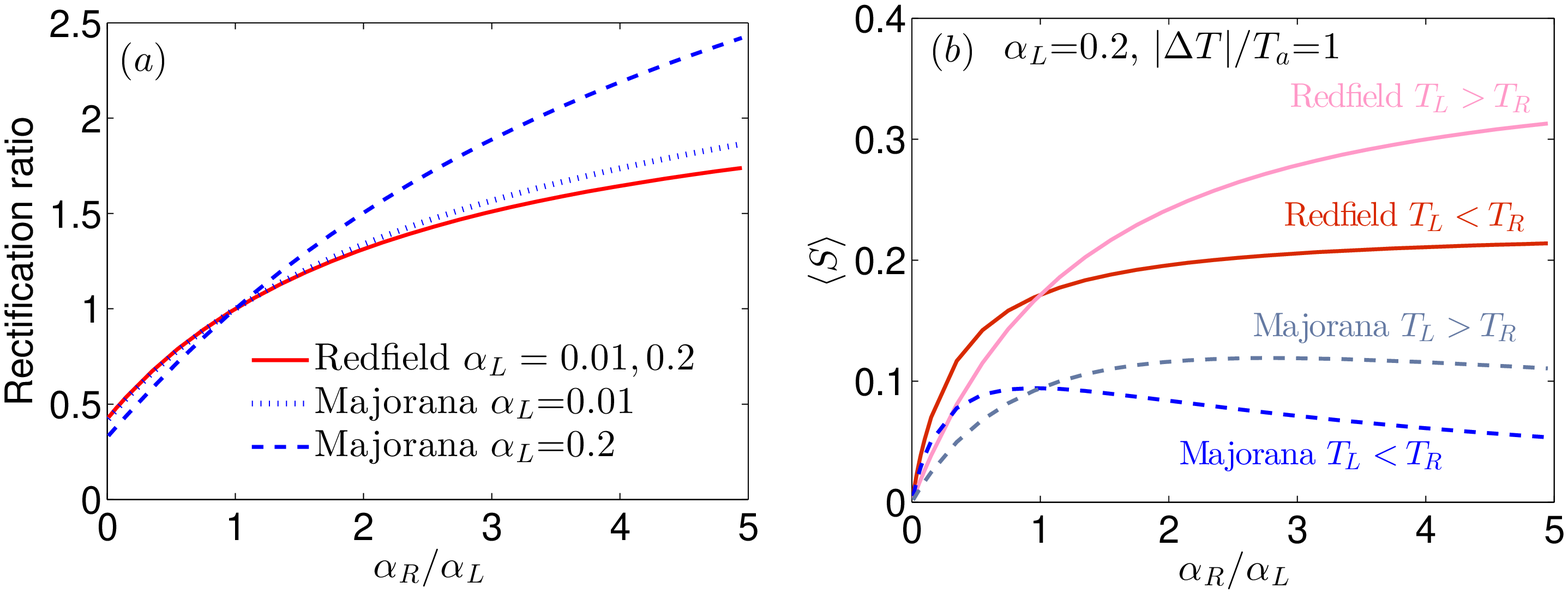}
\caption{ 
Thermal diode effect. (a) Rectification ratio, $R= \langle I(T_L=1.5,T_R=0.5)\rangle| \langle I(T_L=0.5,T_R=1.5)\rangle|$,
as a function of the asymmetry in the system-bath coupling, $\alpha_R/\alpha_L$,
while fixing $\alpha_L$.
(b) Noise $\langle S \rangle$ for forward and backward operations 
as a function of the junction asymmetry using $\alpha_L=0.2$.
Parameters are $\Delta=1$, $\Delta T= T_a$, $(T_L,T_R)$=(1.5, 0.5) and (0.5, 1.5),
$\omega_c=10\Delta$.}
\label{Fig3}
\end{figure}

In Figs. \ref{Fig1}-\ref{Fig3}, we present simulations demonstrating the behavior of the heat current $\langle I \rangle$ 
and the second cumulant $\langle S\rangle $, based on Eq. (\ref{eq:CGF-central}),
as a function of the system-bath coupling, averaged temperature, and temperature difference.
We focus on the following questions regarding the operation of the NESB nanojunction:

(i) How are the current and noise influenced by the system-bath coupling strength? (Fig. \ref{Fig1} and \ref{Fig3}).
(ii) What are the signatures of operation far from equilibrium, as opposed to the linear response regime? 
  (Fig. \ref{Fig1} and \ref{Fig3})
(iii) What is the temperature dependence of the heat current? (Fig. \ref{Fig2})
(iv) Thermal diode effect: Can we enhance this effect if we go beyond the weak spin-bath coupling? (\ref{Fig3})
(v) What is the relation between the Majorana-based treatment and other techniques? (Figs. \ref{Fig1}-\ref{Fig3}).

Fig. \ref{Fig1} displays the current and the noise as obtained from Eqs.~(\ref{cur-noise}),
as well as the weak coupling (Redfield) limit \cite{segal-Nicolin,nazim}, and the NIBA approximation \cite{SegalPRE14,nazim}.
We use an Ohmic spectral function for the baths with an exponential cutoff,  
$\Gamma_{\nu}(\omega)= \pi \alpha_{\nu} \omega e^{-\omega/\omega_c}$.  
%
In accord with previous results (for the heat current \cite{nazim}), we find that Redfield equation dramatically overestimates
the current and the noise in comparison to the (more accurate) Majorana and NIBA results.
Majorana treatment shows  a saturation of the current and its noise at large $\alpha$, while under NIBA these quantities quickly 
decay beyond $\alpha\sim 0.15$. Since the temperature is rather high, $\Delta =T_a$, with $T_a=(T_L+T_R)/2$, we expect
the NIBA to be rather accurate here \cite{Saito13,nazim,SegalPRE14}.
%
We also confirm in panel (a) that
in linear response (LR), the conductance, $\langle I\rangle_{LR}/\Delta T$, 
is proportional to the thermal noise in the junction, in accord with the Green-Kubo relation,
\bea
\langle S\rangle_{eq} = 2T_a^2 \langle I\rangle_{LR}/\Delta T.
\eea
Far from equilibrium [see panel (b)], we obviously observe violations of the above relation. However,
it is interesting to note that the current and noise still
follow a similar functional form within the three different methods.

Fig. \ref{Fig2} displays the temperature dependence of the current and the noise. We study both
the NESB model and a fully harmonic junction, Eq. (\ref{eq:IHO}) and (\ref{eq:SHO}), and
make the following observations:
(i) Comparing the current in the HO and NESB nanojunctions,  
anharmonicity, as realized here by the spin, leads to the suppression of the heat current.
(ii) At weak coupling, $\alpha_{\nu}=0.01$, see panels (a1)-(b1), the Majorana and Redfield approaches for the NESB model 
agree. 
(iii) At intermediate coupling, $\alpha_{\nu}=0.2$, see panels (a2)-(b2), Redfield formalism leads to (nonphysical) high currents, even beyond
the harmonic limit---at low temperatures. 
(iv) At high temperatures and intermediate coupling, Majorana calculations show (a weak) decay of the current with
temperature, see panel (a2), an effect expected to show up in anharmonic nanojunctions \cite{reviewSA}.

Next, we discuss the operation of the NESB as a heat diode, as suggested in Ref. \cite{segal-PRL}.
To materialize this effect, it is necessary to (i) include anharmonic interactions, and (ii) introduce a spatial asymmetry \cite{WuPRL}.
The NESB model naturally includes an anharmonic potential. We break here the left-right symmetry by using different
coupling strengths at the contacts, $\alpha_L\neq \alpha_R$.
In Fig. \ref{Fig3}, we analyze the ratio between the forward and backward currents as we switch
the temperatures of the two baths, $R\equiv |\langle I(T_L,T_R)\rangle|/|\langle I(T_R,T_L)\rangle|$.
We set $\alpha_L$=0.01, 0.2, and modify $\alpha_R$ over a broad range of values.

Based on Eq. (\ref{eq:ISBweak}), we can readily confirm that 
under the Redfield formalism the rectification ratio $R$ {\it does not} depend on the absolute value of 
$\alpha$ (given the linearity of the current with $\alpha$), only on the ratio  $\alpha_R/\alpha_L$. 
In contrast, the Majorana treatment, which goes beyond weak coupling,
reveals that the diode effect is enhanced as we increase the coupling strength itself. This result points out to the crucial
role of many-body interactions in realizing the diode function. 

\section{Conclusions}
\label{Summ}

We have studied the statistics of energy transfer in the nonequilibrium spin-boson model.
By combining Majorana fermion representation for the spin operators with the Schwinger-Keldysh Green's function approach, 
we were able to derive an analytical expression for the CGF of the model. 
This function, which we confirmed here to satisfy the fluctuation symmetry for heat exchange,
hands over the complete information over the energy statistics in the steady state limit. 
Our approach goes beyond the weak-coupling (Redfield) and the co-tunnelling limits.
Surprisingly, the CGF of the NESB model has a similar structure as in the harmonic oscillator junction, besides
sign differences and the appearance of a temperature-dependent transmission function---in the NESB model.
These differences reflect on the nonlinear nature of the spin-boson system. 

We have presented numerical examples for the heat current and its noise, and compared our results to previously-developed quantum master equation
approaches, namely Redfield and the NIBA. We have further demonstrated that a heat diode becomes more effective as we increase
the system-bath coupling.
Additional improvements to the Majorana formulation presented here 
could be made, e.g., by developing a polaron-transformed Majorana fermion-NEGF approach  \cite{Wu16}.
Future work will be focused on simulating counting statistics in the NESB model beyond perturbative approaches \cite{Weiss}.

\acknowledgments
The work of DS and BKA was supported by an NSERC Discovery Grant, 
the Canada Research Chair program, and the CQIQC at the University of Toronto. 


\renewcommand{\theequation}{A\arabic{equation}}
\setcounter{equation}{0}  
\section*{Appendix A: Derivation of the cumulant generating function within an NEGF approach}  

Our goal is to evaluate the generalized current, Eq. (\ref{gen-current}). It is  given in terms of the (dressed) 
lesser $\tilde{\Pi}_{xx}^{<}(t,t')$ = 
$-i\langle \sigma_x (t') \sigma_x(t)\rangle_{\xi}$ and greater 
 $\tilde{\Pi}_{xx}^{>}(t,t')=-i
\langle \sigma_x (t) \sigma_x(t')\rangle_{\xi}$ correlators.
Keeping in mind the nonequilibrium setup, we introduce the $\xi$-dependent contour-ordered Green's function for the $\sigma_x$ component,
\bea
\tilde{\Pi}_{xx}(\tau,\tau')&=& -i \langle T_c \sigma_x(\tau) \sigma_x(\tau') \rangle_{\xi} \nonumber \\
&= & \begin{bmatrix}\tilde{\Pi}_{xx}^{t}\left(t,t'\right) &
\tilde{\Pi}_{xx}^{<}\left(t,t'\right)\\
\tilde{\Pi}_{xx}^{>}\left(t,t'\right) &
\tilde{\Pi}_{xx}^{\bar{t}}\left(t,t'\right)
\end{bmatrix}.
\eea
Recall that $\langle \cdots \rangle_{\xi}$ means that operators are evolving with the dressed Hamiltonian of 
Eq.~(\ref{eq:mod-Hamiltonian}). 
Here $\tau,\tau'$ are the contour times. When projecting to real time $(t,t')$,  
we receive  four different terms, namely,  
time-ordered ($t$), anti-time ordered ($\bar{t})$, lesser $(<)$ and greater $(>)$ Green's functions. 

To evaluate the greater and lesser components, we use the Majorana fermion representation of spin operators,
Eqs. (\ref{eq:maj1})-(\ref{eq:maj4}). We  identify our objective of interest by
$-i \langle T_c \left(f(\tau)+f^{\dagger}(\tau)\right) \left(f(\tau')+f^{\dagger}(\tau')\right) \rangle_{\xi}$.
We define Green's function for the Dirac $f$ fermion in the Bogolyubov-Nambu representation i.e., 
$\Psi \equiv (f, f^{\dagger})^T$ and $\Psi^{\dagger} \equiv (f^{\dagger}, f)$, 
and write $\hat{\tilde{{G}}}_{\Psi}(\tau,\tau')= -i \langle T_c \Psi(\tau) \Psi(\tau') \rangle_{\xi} $. 
The symbol hat in $\hat{\tilde {{G}}}_{\Psi}(\tau,\tau')$ represents a $2 \times 2$ matrix in the contour space ($4 \times 4$ in real time) i.e.,
\be
\hat{\tilde {{G}}}_{\Psi}(\tau,\tau')= 
\begin{bmatrix}
-i \langle T_c f(\tau) f^{\dagger}(\tau')\rangle_{\xi}  & \, -i \langle T_c  f(\tau) f(\tau')\rangle_{\xi} \\
-i \langle T_c  f^{\dagger}(\tau) f^{\dagger}(\tau')\rangle_{\xi}  &\,\, -i \langle T_c  f^{\dagger}(\tau) f(\tau')\rangle_{\xi} 
\end{bmatrix}
\ee
and in real time 
\be
\hat{\tilde {{G}}}_{\Psi}(t,t')= 
\begin{bmatrix}
\tilde{G}_{\psi}^{t}(t,t')  & \,  \tilde{G}_{\psi}^{<}(t,t') \\
 \tilde{G}_{\psi}^{>}(t,t')  &\, \tilde{G}_{\psi}^{\bar{t}}(t,t').
\end{bmatrix}.
\ee
Each component comprises a $2 \times 2$ matrix.
Then the $\tilde{\Pi}_{xx}^{<,>}(\omega)$ components can be alternatively expressed as
\bea
\tilde{\Pi}_{xx}^{<}(\omega) &=& -\left(\begin{array}{cc} 1 & 1 \end{array}\right ) {\tilde{G}}_{\Psi}^{<}(\omega) \left(\begin{array}{c}1 \\1 \end{array}\right), \nonumber \\
\tilde{\Pi}_{xx}^{>}(\omega) &=& \left(\begin{array}{cc} 1 & 1 \end{array}\right ) {\tilde{G}}_{\Psi}^{>}(\omega) \left(\begin{array}{c}1 \\1 \end{array}\right).
\label{eq:tildePi}
\eea
%
%
We next construct a Dyson (kinetic) equation for $\hat{\tilde {{G}}}_{\Psi}(\tau,\tau')$ following the dressed Hamiltonian in 
Eq.~(\ref{eq:mod-Hamiltonian}) treating the nonlinear part of the Hamiltonian, $\left(f^{\dagger}-f\right)\eta_z(B_L^p+B_R)$, as  a perturbation,
\be
\hat{\tilde {{G}}}_{\Psi}(\tau,\tau') = \hat{{G}}_{\Psi,0}(\tau,\tau') + \int d\tau_1 \int d\tau_2 \hat{G}_{\Psi,0}(\tau,\tau_1) \, \hat{\tilde{\Sigma}}_{\Psi}(\tau_1,\tau_2)\, \hat{\tilde {{G}}}_{\Psi}(\tau_2,\tau').
\label{Dyson_spin_1}
\ee
The Green's functions of the ordinary fermion $g$ and the reservoirs operators are calculated to the lowest (noninteracting) order.
We thus write the contour ordered version of the self-energy $\hat{\tilde{\Sigma}}_{\Psi}(\tau_1,\tau_2)$ as
\be
\hat{\tilde{\Sigma}}_{\Psi}(\tau,\tau')=i \hat{\lambda} \Big(\tilde{\Sigma}_L(\tau,\tau')+\Sigma_R(\tau,\tau')\Big)\, G_{\eta}(\tau,\tau').
\label{non-self}
\ee 
$\hat{{G}}_{\Psi,0}(\tau,\tau')$ is the Green's function corresponding to the the noninteracting part of the Hamiltonian. It
satisfies the following differential equation in contour time 
\be
(i \partial_{\tau} \hat{I}- \Delta \hat{\sigma}_z) \hat{{G}}_{\Psi,0}(\tau,\tau')= \hat{\delta}(\tau-\tau'),
\ee
where $\hat{I}$ is $2\times2$ identity matrix. In frequency domain, we obtain the solution
$\hat G_{\Psi,0}^{-1}(\omega)= {\rm diag} (\omega-\Delta, \omega+\Delta, -\omega+\Delta, -\omega-\Delta)$. 

In Eq.~(\ref{non-self}), ${G}_{\eta}(\tau,\tau') = -i \langle T_c \eta_z(\tau) \eta_z(\tau') \rangle$ 
is the Green's function involving the $z$-th component of the Majorana fermion,  
$\hat{\lambda}$ is the Nambu matrix
\be
\hat{\lambda}= 
\begin{bmatrix}
\,\,1 & \,-1 \\
-1 &\,\,\,\,1 
\end{bmatrix}
\ee
and $\tilde{\Sigma}_{L}, {\Sigma}_{R}$ are the bare Green's functions for the Bosonic baths,
\bea
\tilde{\Sigma}_{L}(\tau,\tau')= -i \langle \tilde{B}_L(\tau) \, \tilde{B}_L(\tau') \rangle, \nonumber \\
{\Sigma}_{R}(\tau,\tau')= -i \langle {B}_R(\tau) \, {B}_R(\tau') \rangle,
\eea
Recall that the operators of the left reservoirs are dressed by the additional $\xi$ dependence, 
i.e., $\tilde{B}_L(\tau) = B_L^{-\xi/2}(t)$ ($B_L^{+\xi/2}(t)$), when $\tau$ is on the upper (lower) branch.  
Given the perturbative nature of our treatment, the self-energy contribution from the baths is additive.
%
%

To the lowest non-zero order, various components of the self-energy can be obtained analytically. 
Invoking the steady state limit, we write down these components in  frequency domain, 
given by the convolution of the Green's functions for the baths and the Majorana fermions.
Following Eq.~(\ref{non-self}), we get
\bea
\tilde{\Sigma}_{\Psi}^{>,<}(\omega)&=& {i} \hat{\lambda} \int \frac{d\omega'}{2 \pi} \tilde{\Sigma}^{>,<}_{X}(\omega-\omega') \,G^{>,<}_{\eta}(\omega'),
\eea
where we use the notation $\tilde{\Sigma}^{>,<}_X = \tilde{\Sigma}^{>,<}_L + {\Sigma}^{>,<}_R$. 
This expression can be further simplified by using symmetry relations, as follows. 
The sum and difference of the lesser and greater components are given by 
%
%
\bea
\tilde{\Sigma}_{\Psi}^{K}(\omega)=  \frac{i \hat{\lambda}}{2} \int \frac{d\omega'}{2\pi} \Big[ (\tilde{\Sigma}^{>}_X-\tilde{\Sigma}^{<}_X)(\omega+\omega')(G_{\eta}^r(\omega')- G_{\eta}^a(\omega')) - \tilde{\Sigma}_X^{K}(\omega+\omega') {G}^{K}_{\eta}(\omega')\Big] \,\,.
\eea
and 
\bea 
\tilde{\Sigma}_{\Psi}^{>}(\omega) - \tilde{\Sigma}_{\Psi}^{<}(\omega) &=& \frac{i \hat{\lambda}}{2}  \int \frac{d\omega'}{2\pi} \Big[\tilde{\Sigma}^{K}_X(\omega+\omega') (G_{\eta}^r(\omega')\!-\! G_{\eta}^a(\omega')) -(\tilde{\Sigma}^{>}_X (\omega+\omega')\!-\! \tilde{\Sigma}^{<}_X (\omega+\omega')) G_{\eta}^K(\omega')\Big],
\nonumber \\
\eea
Here $K$ is the Keldysh component, the sum of lesser and greater Green's functions.
The spectral function of the ordinary fermion, $\Gamma_{\eta}(\omega)= \frac{i}{2} \big(G^r_{\eta}(\omega)-G^a_{\eta}(\omega))= 2 \pi \delta(\omega)$, 
satisfies the regular sum rule $\int \frac{d\omega}{2 \pi} \Gamma_{\eta}(\omega) =1$. 
We also use the effective fluctuation-dissipation relation i.e., $G^{K}_{\eta}(\omega) = h_{\eta} (\omega) \big(G^r_{\eta}(\omega)-G^a_{\eta}(\omega)) \propto h_{\eta} (\omega) \delta(\omega) = 0$ where $h_{\eta}(\omega)=\tanh(\frac{\beta_L \hbar \omega}{2}) +\tanh(\frac{\beta_R \hbar \omega}{2})$.  
Putting these pieces together, we obtain simplified expressions for the self-energy components, 
expressed solely in terms of the reservoirs' self-energies, 
\be
\tilde {\Sigma}_{\Psi}^{>}(\omega) = \hat{\lambda}\,  \tilde{\Sigma}_X^{>}(\omega), \quad \tilde{\Sigma}_{\Psi}^{<}(\omega) = -\hat{\lambda}\, \tilde {\Sigma}_X^{<}(\omega).
\label{self-1}
\ee
We next look at the time ordered and anti-time ordered components. These terms are $\xi$-independent, and they satisfy the following relations 
\bea
{\Sigma}_{\Psi}^{t}(\omega) + {\Sigma}_{\Psi}^{\bar{t}}(\omega) &=& {\Sigma}_{\Psi}^{>}(\omega) + {\Sigma}_{\Psi}^{<}(\omega) =\hat{\lambda}\, (\Sigma_X^r(\omega)-\Sigma_X^a(\omega)) \\
{\Sigma}_{\Psi}^{t}(\omega) -{\Sigma}_{\Psi}^{\bar{t}}(\omega) &=& \frac{i \hat{\lambda}}{2} \int \frac{d\omega}{2\pi} \Sigma_X^K(\omega) (G_{\eta}^r(\omega) + G_{\eta}^a(\omega))=0.
\eea
To derive the last expression, we ignore the lamb shift part. We therefore find that
\bea
{\Sigma}_{\Psi}^{t}(\omega) &=&{\Sigma}_{\Psi}^{\bar t}(\omega) = - \frac{i \hat{\lambda}}{2} \Big(\Gamma_L(\omega) + \Gamma_R(\omega)\Big),
\label{self-2}
\eea
where $\Gamma_{\nu}(\omega) = i (\Sigma_{\nu}^r(\omega)-\Sigma_{\nu}^a(\omega))$ describes 
the reservoir-subsystem coupling energy.
Therefore, to the lowest order in perturbation theory, 
the self-energy $\hat{\tilde{\Sigma}}_{\Psi}$ is fully determined by the reservoir's Green's functions as given by Eq.~(\ref{self-1}) 
and Eq.~(\ref{self-2}). 

We can now solve Eq.~(\ref{Dyson_spin_1}), by projecting it to real time then Fourier transforming it,
\be
\bar{\tilde{G}}_{\Psi}^{-1}(\omega) ={\bar {G}}_{\Psi,0}^{-1}(\omega) - {\bar{\tilde{\Sigma}}}_{\Psi}(\omega) =
\\
{\begin{bmatrix} 
\omega-\Delta + \frac{i}{2} \Gamma(\omega) &  -\frac{i}{2} \Gamma(\omega) & \tilde{\Sigma}_X^<(\omega) & -\tilde{\Sigma}_X^<(\omega) \\
-\frac{i}{2} \Gamma(\omega)  & \omega+\Delta+\frac{i}{2} \Gamma(\omega)  & -\tilde{\Sigma}_X^<(\omega) & \tilde{\Sigma}_X^<(\omega) \\
\tilde{\Sigma}_X^>(\omega) & -\tilde{\Sigma}_X^>(\omega) & \omega-\Delta-\frac{i}{2} \Gamma(\omega)  & 
\frac{i}{2} \Gamma(\omega)  \\
-\tilde{\Sigma}_X^>(\omega) & \tilde{\Sigma}_X^>(\omega) & \frac{i}{2} \Gamma(\omega)  & \omega+\Delta -\frac{i}{2} \Gamma(\omega) 
\end{bmatrix}}
\ee
Here, the symbol bar represents $\bar{A} = {\hat{\hat {\sigma}}}_z A$  with ${\hat{\hat {\sigma_z}}} = {\rm diag} (1, 1, -1, -1)$, introduced so as to take into account the appropriate signs for upper and lower branches of the contour. 
$\Gamma(\omega)= \Gamma_L(\omega)+ \Gamma_R(\omega)$. 
Inverting this matrix we obtain the lesser component 
\begin{equation}
\tilde{G}_{\Psi}^{<}(\omega) = \frac{1}{{\cal D}(\omega,\xi)}
{\begin{bmatrix}
(\omega+ \Delta)^2 \, \tilde{\Sigma}_{X}^{<}(\omega) & -(\omega^2-\Delta^2) \, \tilde{\Sigma}_{X}^{<}(\omega)\\
-(\omega^2-\Delta^2 )\, \tilde{\Sigma}_{X}^{<}(\omega) & (\omega- \Delta)^2 \, \tilde{\Sigma}_{X}^{<}(\omega)
\end{bmatrix}},
\end{equation}
and the greater component 
\begin{equation}
\tilde{G}_{\Psi}^{>}(\omega) = -\frac{1}{\cal D(\omega,\xi)}
{\begin{bmatrix}
(\omega \!+\! \Delta)^2 \, \tilde{\Sigma}_{X}^{>}(\omega) & \,\,  -(\omega^2 \!-\! \Delta^2)\, \tilde{\Sigma}_{X}^{>}(\omega)\\
-(\omega^2 \!-\! \Delta^2)\, \tilde{\Sigma}_{X}^{>}(\omega) & \,\,(\omega\!-\! \Delta)^2 \,\tilde{\Sigma}_{X}^{>}(\omega)
\end{bmatrix}}.
\end{equation}
${\cal D(\omega,\xi)}$ is the determinant of the matrix, given as
\bea
{\cal D}(\omega,\xi)&=& (\omega^2 -\Delta^2)^2 + \omega^2 \Big[\Big(\Gamma_L(\omega) (1 + 2{n}_L(\omega)) + \Gamma_R(\omega) 
(1+2 {n}_R(\omega))\Big)^2 + 4 \Gamma_L(\omega) \Gamma_R(\omega) \big\{ n_L(\omega)\bar{n}_R(\omega) \nonumber \\
&&( e^{i \xi \hbar \omega}\!-\!1) + n_R(\omega)\bar{n}_L(\omega) (e^{-i \xi \hbar \omega}\!-\!1)\big\} \Big].
\eea
with $\bar{n}_{\nu}(\omega)= 1 + n_{\nu}(\omega)$.
Using Eq.  (\ref{eq:tildePi}), the lesser and greater components of spin-spin correlation functions are finally obtained as 
\bea
\tilde{\Pi}_{xx}^{<}(\omega) 
&=& -\frac{4 i \Delta^2}{{\cal D}(\omega,\xi)} \Big( \Gamma_L(\omega)  n_L(\omega) e^{i \xi \hbar \omega} + \Gamma_R(\omega) n_R(\omega) \Big) \nonumber \\
\tilde{\Pi}_{xx}^{>}(\omega)
&=& -\frac{4 i \Delta^2}{{\cal D}(\omega,\xi)} \Big( \Gamma_L(\omega)  \bar{n}_L(\omega) e^{-i \xi \hbar \omega} + \Gamma_R(\omega) \bar{n}_R(\omega) \Big).
\eea
Substituting these expressions into the generalized current expression,  Eq.~(\ref{gen-current}), we receive 
\be
{\cal I}(\xi)= \int_{-\infty}^{\infty} \frac{d\omega}{4\pi} \frac{\Delta^2}{\omega^2} \, \frac{1}{{\cal D}(\omega,\xi)} \, \frac{\partial}{\partial(i \xi)} \Big[{\cal D}(\omega,\xi)\Big].
\ee
Manipulating it as ${\cal G}(\xi) = \int^{\xi}_{0} {\cal I}(\xi') d\xi'$, we get
\bea
{\cal G}(\xi) = \int_{-\infty}^{\infty} \frac{d\omega}{4\pi} \frac{\Delta^2}{\omega^2} 
\ln\left[\frac{ {\cal D}(\omega,\xi)}{  {\cal D}(\omega,\xi=0)}\right],
\eea  
which we organize into our main result, Eq.~(\ref{eq:CGF-central}).


\begin{thebibliography}{0}

\bibitem{Weiss-book} 
U. Weiss, {\it Quantum Dissipative Systems} (World Scientific, Singapore, 1999).

\bibitem{Leggett-spin} 
A. J. Leggett, S. Chakravarty, A. T. Dorsey, M. P. A. Fisher, A. Garg, and W. Zwerger, 
Rev. Mod. Phys. {\bf 59}, 1 (1987).

\bibitem{Nitzan-book} 
A. Nitzan, {\it Chemical Dynamics in Condensed Phases} (Oxford University Press, New York, 2006).

\bibitem{Hur-spin} 
K. Le Hur, {\it Understanding Quantum Phase Transitions}, edited by L. D. Carr (Taylor and Francis, Boca Raton, 2010).

%


\bibitem{segal-PRL} 
D. Segal  and  A. Nitzan,  Phys. Rev. Lett {\bf 94},  034301 (2005).

\bibitem{segal-QME} D. Segal, 
Phys. Rev. B {\bf 73}, 205415 (2006).


\bibitem{segal-JCP05} D. Segal and  A. Nitzan, J. Chem. Phys. {\bf 122}, 194704 (2005).

\bibitem{WuPRL}
L.-A. Wu and D. Segal,
Phys. Rev. Lett. {\bf 102}, 095503 (2009).

\bibitem{ClaireQME}
L.-A. Wu, C. X. Yu, and D. Segal, 
Phys. Rev. E {\bf 80}, 041103 (2009).


\bibitem{Saito13} K. Saito and T. Kato, 
Phys. Rev. Lett. {\bf 111}, 214301 (2013).


\bibitem{nazim} N. Boudjada and D. Segal, J. Phys. Chem. A, {\bf 118}, 11323 (2014).


\bibitem{reviewSA} D. Segal and B. K. Agarwalla., Annu. Rev. Phys. Chem. {\bf 67}, 185 (2016).


\bibitem{review-wang}
J.-S. Wang,  J. Wang, J. T. Lu, 
Eur. Phys. J. B {\bf 62}, 381 (2008).

\bibitem{review-li} 
N. Li {\it et al.}
Rev. Mod. Phys. {\bf 84}, 1045 (2012).


\bibitem{segal-pump} 
D. Segal and A. Nitzan, 
Phys. Rev. E {\bf 73}, 026109 (2006).

\bibitem{segal-stochastic} 
D. Segal, 
Phys. Rev. Lett. {\bf 101}, 260601 (2008).

\bibitem{berry-1} J. Ren, P. H\"anggi, and B. Li,  Phys. Rev. Lett. {\bf 104}, 170601 (2010).


\bibitem{berry-2}  
T. Chen, B. X. Wang, and J. Ren, 
Phys. Rev. B {\bf 87}, 144303 (2013).


%
%
%

%

%
%
%

\bibitem{siminePCCP}
L. Simine and D. Segal,
Phys. Chem. Chem. Phys. {\bf 14}, 13820 (2012).

\bibitem{simine-an}
L. Simine and D. Segal,
J. Chem. Phys. {\bf 141}, 014704 (2014).


\bibitem{bijay-segal-fcs} 
B. K. Agarwalla, J.-H. Jiang, and D. Segal, 
Phys. Rev. B {\bf 92}, 245418 (2015). 



\bibitem{book-ME} 
J. C. Cuevas and E. Scheer, 
{\it Molecular Electronics: An Introduction to Theory and Experiment}, World Scientific, Singapore , 2010.


\bibitem{Galperin} 
M. Galperin, M. A. Ratner, A. Nitzan, 
J. Phys. Condens. Matter {\bf 19}, 103201 (2007).


\bibitem{beil}
B. K. Agarwalla, J.-H. Jiang and D. Segal,
Beilstein J. Nanotechnol. {\bf 6}, 2129 (2015).

\bibitem{Ojanen-QME} 
T. Ruokola and T. Ojanen, 
Phys. Rev. B {\bf 83}, 045417 (2011).


\bibitem{Juzar}
J. Thingna, H. Zhou, J. S. Wang,
J. Chem. Phys. {\bf 141}, 194101 (2014).
%
\bibitem{renJ}
J. Ren, P. H\"anggi, and B. Li, 
Phys. Rev. Lett. {\bf 104}, 170601 (2010).



\bibitem{Ren} 
C. Wang, R. Jie. J. Cao, 
Sci. Rep. {\bf 5}, 11787 (2015).


\bibitem{segal-Nicolin} 
L. Nicolin and  D. Segal,  
J. Chem. Phys. {\bf 135}, 164106 (2011).




\bibitem{ThossNEGF}
K. A. Velizhanin, M. Thoss, and H. Wang,
J. Chem. Phys.  {\bf 133}, 084503 (2010).


\bibitem{yang-EPL} 
Y. Yang and C. Q. Wu, 
Euro. Phys. Lett. {\bf 107}, 30003 (2014).


\bibitem{num-multi} 
K. A. Velizhanin, H. Wang, and M. Thoss, 
Chem. Phys. Lett. {\bf 460}, 325 (2008).

\bibitem{num-infi} 
D. Segal, Phys. Rev. B {\bf 87}, 195436 (2013).


\bibitem{tanimura}
A. Kato and Y. Tanimura,
J. Chem. Phys, {\bf 143}, 064107 (2015).

\bibitem{brandes}
J. Cerrillo, M. Buser, T. Brandes,
arXiv:1606.05074.




\bibitem{Esposito-review} 
M. Esposito, U. Harbola, and S. Mukamel, Rev. Mod. Phys. {\bf 81}, 1665 (2009).

\bibitem{Hanggi-review} 
M. Campisi, P. H\"anggi, and P. Talkner, Rev. Mod. Phys. {\bf 83}, 771 (2011).


\bibitem{FCS-strong} L. Nicolin and D. Segal, 
Phys. Rev. B {\bf 84}, 161414 (2011).




%
\bibitem{Rammer-NEGF} 
J. Rammer, {\it Quantum Field Theory of Non-Equilibrium States} (Cambridge University Press, 2007). 
%
%
\bibitem{bijay-wang-review} 
J.-S. Wang, B. K. Agarwalla, H. Li, and J. Thingna, Front. Physics {\bf 9}, 673 (2014).



\bibitem{SegalPRE14}
D. Segal,
Phys. Rev. E {\bf 90}, 012148 (2014).

\bibitem{spin1} 
A. Shnirman  and Y. Makhlin, 
Phys. Rev. Lett. {\bf 91}, 207204 (2003).

\bibitem{spin2} 
W. Mao, P. Coleman, C. Hooley, and D. Langreth. 
Phys. Rev. Lett. {\bf 91}, 207203 (2003).




\bibitem{Gogolin} 
A. O. Gogolin and A. Komnik, Phys. Rev. B {\bf 73},  195301 (2006).

\bibitem{Schwinger} 
J. Schwinger, Brownian motion of a quantum oscillator, J. Math. Phys., {\bf 2}, 407 (1961).

\bibitem{Kadanoff} 
L. P. Kadanoff and G. Baym, {\it Quantum Statistical Mechanics}, Benjamin/Cummings, 1962

\bibitem{Keldysh} 
L. V. Keldysh, Diagram technique for nonequilibrium processes, Sov. Phys. JETP,  {\bf 20}, 1018 (1965).

\bibitem{book1} P. Danielewicz,
 {\it Quantum theory of nonequilibrium processes (I)}, Ann. Phys.,  {\bf 152}, 239 (1984).

\bibitem{bijay-Euro} 
H. Li,  B. K. Agarwalla,  B. Li, and  J. -S. Wang, Eur. Phys. J. B {\bf 86},  1 (2013).

\bibitem {meir-wingreen} Y. Meir and N. Wingreen, 
Phys. Rev. Lett. {\bf 68}, 2512 (1992).

\bibitem{GC-sym} G. Gallavotti and E. G. D. Cohen, 
Phys. Rev. Lett. {\bf 74}, 2694 (1995).


\bibitem{BO}
L.-A. Wu and D. Segal,
Phys. Rev. E {\bf 83}, 051114 (2011).

\bibitem{saito-fcs} 
K. Saito and A. Dhar, 
Phys. Rev. Lett. {\bf 99}, 180601 (2007).

\bibitem{bijay-PRE-fcs} 
B. K. Agarwalla, B. Li, and  J.-S. Wang, 
Phys. Rev. E {\bf 85}, 051142 (2012).




\bibitem{Andrei}
Y. Vinkler-Aviv, A. Schiller, and N. Andrei,
Phys. Rev. B {\bf 89}, 024307 (2014).


\bibitem{Ed-bound}
E. Taylor and D. Segal,
Phys. Rev. Lett. {\bf 114}, 220401 (2015).

\bibitem{Weiss} 
M. Carrega, P. Solinas, A. Braggio, M. Sassetti, and U. Weiss, 
New J. Phys. {\bf 17}, 045030 (2015).


\bibitem{Wu16}
J. Liu, H. Xu, and C.-Q. Wu,  
http://dx.doi.org/10.1016/j.chemphys.2016.07.003 
Chem. Phys. {\bf 2016}


\end{thebibliography}
\end{document}